\documentclass[12pt]{article}
\usepackage{graphicx}
\usepackage{bm}
\def\gpp{p\!\!\!/}
\def\gpq{q\!\!\!/}
\begin{document}
\newcommand{\cd}{\makebox[0.08cm]{$\cdot$}}
\centerline{\bf $qq\bar {q} \to qq\bar q$ and
$q\bar {q}\bar {q} \to q\bar {q}\bar q$ Elastic Scatterings}
\centerline{\bf and Thermalization of Quark Matter and Antiquark Matter}
\vskip 18pt
\centerline{Xiao-Ming X${\rm u}^{\rm a}$,
Cheng-Cheng M${\rm a}^{\rm a}$, An-Qian Che${\rm n}^{\rm b}$,
H.J. Webe${\rm r}^{\rm c}$}
\vskip 14pt
\centerline{$^{\rm a}$Department of Physics, Shanghai University, Baoshan,
Shanghai 200444, China}
\centerline{$^{\rm b}$Department of Communication, Shanghai University,
Baoshan, Shanghai 200444, China}
\centerline{$^{\rm c}$Department of Physics, University of Virginia,
Charlottesville, VA 22904, U.S.A.}

\begin{abstract}
\baselineskip=14pt
Thermalization of quark matter and antiquark matter is studied with
quark-quark-antiquark as well as quark-antiquark-antiquark elastic
scatterings. Squared amplitudes of  $qq\bar {q} \to qq\bar q$ and
$q\bar {q}\bar {q} \to q\bar {q}\bar q$ at order $\alpha_s^4$ are derived in
perturbative QCD. Solved by a new technique,
solutions of transport equations with the squared amplitudes
indicate that the scatterings  $qq\bar {q} \to qq\bar q$ and
$q\bar {q}\bar {q} \to q\bar {q}\bar q$ shorten the thermalization time of
quark matter and antiquark matter. It is emphasized that three-parton and
other multi-parton scatterings become important at the high parton number
density achieved in RHIC Au-Au collisions.
\end{abstract}
\leftline{PACS codes: 24.85.+p; 12.38.Mh; 25.75.Nq}
\leftline{Keywords: quark and antiquark matter; quark-quark-antiquark elastic
scattering;}
\leftline{~~~~~~~~~~~~~~~transport equation; thermalization}

\newpage
\leftline{\bf 1. Introduction}
\vspace{0.5cm}
Establishing a thermal state is critical in the formation of a quark-gluon
plasma [1,2]. Initial high-energy heavy-ion collisions create deconfined matter
which is not in thermal and chemical equilibrium. The problem of quark-gluon
matter thermalization was studied early in parton dynamics in Refs. [3,4].
The 2-to-2 and 2-to-3 gluon scatterings produce a gluon-matter thermalization
time greater than 1 fm/c [5-7]. In accounting for the elliptic flow of hadrons
observed in Au-Au collisions at the Relativistic Heavy Ion Collider (RHIC),
recent hydrodynamic calculations come to the conclusion that the 
thermalization of quark-gluon matter is finished within the time of 1 fm/c 
[8-11]. The thermalization process is fascinating and needs to be understood 
urgently. We have proposed 3-to-3 as well as 2-to-2 quark elastic scatterings 
to study the thermalization of quark matter [12]. But a thermalization time 
of about 1.8 fm/$c$ is obtained from the mechanism. How rapid thermalization
of quark matter comes about from quark scatterings is still a difficult and
complicated task
to explain. In this letter we include quark-quark-antiquark 
and quark-antiquark-antiquark elastic 
scatterings in an effort to understand better the thermalization of both 
quark and antiquark matter independent of gluon matter.

We present four new aspects. The first is to recognize the importance of
multi-parton scatterings at a high parton number density; the second is
to calculate perturbatively the squared amplitudes of
quark-quark-antiquark elastic scatterings which subsequently give the squared
amplitudes of quark-antiquark-antiquark elastic scatterings; the third is to
solve transport equations with a new technique; the fourth is to provide a new
form for the formula of the interaction range of three-parton scattering. The
squared amplitudes are derived by Fortran codes that implement the QCD 
Feynman rules. The new technique deals with a multi-parton scattering 
in terms of a
sphere that moves with a particle. These four aspects form an integral
part of our study of the thermalization of quark and antiquark matter.

\vspace{0.5cm}
\leftline{\bf 2. Probability of multi-parton scatterings}
\vspace{0.5cm}
The rapid thermalization is related to high gluon number density [13-15]
at which triple-gluon elastic scatterings lead to a short thermalization
time [16]. The importance of three-gluon scatterings is implied and we can thus
speculate on possible or even sizeable contributions of other multi-gluon
scatterings. We therefore give a Monte Carlo test on the occurrence of
multi-parton scatterings from a parton distribution that is anisotropic
in momentum space and inhomogeneous in coordinate space, as created in
initial collisions of two Lorentz-contracted gold nuclei. Such a parton
distribution has been generated by HIJING [17] for central Au-Au collisions at
 $\sqrt {s_{NN}}=200$ GeV and cast into the form [18]
\begin{equation}
f(k_\bot,y,r,z,t)=\frac {1}{16\pi R_A^2}g(k_\bot,y)\frac
{e^{-(z-t\tanh y)^2/2\Delta_k^2}}{\sqrt {2\pi}\Delta_k},
\end{equation}
with
\begin{displaymath}
\Delta_k \approx \frac {2}{k_\bot \cosh y}
\end{displaymath}
and
\begin{displaymath}
g(k_\bot,y)=\frac {(2\pi)^3}{k_\bot \cosh y} \frac {dN}{dyd^2k_\bot},
\end{displaymath}
where the gold nucleus radius is $R_A$=6.4 fm, and $k_\bot$, $y$, $t$, $z$
and $r$ are transverse momentum, rapidity, time, coordinate and radius in the
transverse direction, respectively. One thousand and five hundred gluons are
created from the distribution $f(k_\bot,y,r,z,0.2~{\rm fm}/c)$ by the
rejection method within $-0.3~{\rm fm}<z<0.3~{\rm fm}$ in the longitudinal
direction and $r<R_A$ in the transverse direction. This volume of partons
corresponds to a parton number density of 19.4 ${\rm fm}^{-3}$. At such high
density, we have to examine the occurrence of multi-parton scatterings. We
assume a scattering of two partons when the distance of the two partons is
less than a given interaction range. If $n$ partons are within a sphere of
which the center is at the center of mass of the $n$ partons and of which the
radius equals the given interaction range, a scattering of the $n$ partons
is taken to occur. When the scattering of $n$ partons with a certain value of
$n$ is counted, other multi-parton scatterings are excluded. Therefore, the
maxima of the numbers of 2-parton, 3-parton, 4-parton, 5-parton, 6-parton and
7-parton scatterings at a given time are 750, 500, 375, 300, 250 and 214,
respectively, which give a sum of 2389. The numbers of $n$-parton scatterings
with $n=2,\cdot \cdot \cdot ,7$ at $t=0.2~{\rm fm}/c$ are denoted each by
$m_n$ and plotted in Fig. 1. When the interaction range increases, more
scatterings of not only two but also seven partons happen. All the scattering
numbers saturate at an interaction range of 0.6 fm except for the three-parton
 scattering number which saturates at a smaller interaction range. If we 
consider the practical case when 2-parton scattering, 3-parton scattering, 
etc. may happen at the same time, we need to estimate the probability of 
$n$-parton scattering. A guide for this probability is provided by 
$R_n=m_n/2389$, the ratio of the $n$-parton scattering number obtained above 
to the total number 2389. This ratio is drawn in Fig. 2. For an interaction 
range larger than 0.15 fm, the three-parton scattering becomes important in 
comparison to the two-parton scattering. At an interaction range of 0.62 fm,
which acts as the radius of a unit volume of sphere, the 2-parton scattering 
has the occurrence probability of 30 \%, the 3-parton scattering 20 \%, 
the 4-parton scattering 14.6 \%, and 
even the 7-parton scattering has 7.5 \%. The scattering numbers and the ratios 
shown in Figs. 1 and 2 lead to the high parton number density of 19.4 
${\rm fm}^{-3}$ which implies a substantial occurrence of multi-parton
scatterings.

\vspace{0.5cm}
\leftline{\bf 3. Quark-quark-antiquark elastic scatterings}
\vspace{0.5cm}
In Figs. 3 and 4 ten diagrams are plotted to show the elastic scattering
of quark-quark-antiquark at order $\alpha_{\rm s}^4$. The eight diagrams in
Fig. 3 are related to two successive gluon exchanges and the two diagrams
in Fig. 4 involve a triple gluon coupling. If the two initial or final quarks 
have the same flavor, the exchange of the two quarks generates other diagrams
which are not displayed in Figs. 3 and 4.
For ${\rm uu\bar u} \to {\rm uu\bar {u}}$, we must take into account
38 diagrams of which 28 diagrams contain quark-antiquark annihilation and
creation, and then confront a cumbersome derivation of the squared amplitude 
for the scattering
$q(p_1)+q(p_2)+{\bar q}(-p_3) \to q(p_4)+q(p_5)+{\bar q}(-p_6),$ where
$p_{\rm i}=(E_{\rm i},\vec {p}_{\rm i})$ is the quark four-momentum. 

As an example, the derivation of the squared amplitude of the last diagram in
Fig. 3 is described here. Denote by $q_1$, $g$ and $\lambda$ the momentum, 
color and space-time index of the gluon created by the annihilation of the
initial quark and antiquark, by $q_3$, $h$ and $\sigma$ of the other gluon, 
and by $q_2$ the momentum of the quark propagator, respectively. With the 
Feynman rules given in Ref. [19], the first Fortran code is designed to 
construct the spin- and color-summed squared amplitude 
\begin{eqnarray}
\sum\limits_{\rm {spins,colors}}
\mid {\cal M}_{{\rm F}_{*-}} \mid^2
& = & \frac {{\rm g}_{\rm s}^8}{q_1^4q_2^4q_3^4}
\sum\limits_{\rm {colors}} tr(T_gT_hT_{h^\prime}T_{g^\prime})
tr(T_hT_{h^\prime})tr(T_gT_{g^\prime})
                   \nonumber   \\
& &
tr(\gamma_\lambda \gpq_2 \gamma_\sigma \gpp_1 \gamma_{\sigma^\prime} \gpq_2
\gamma_{\lambda^\prime} \gpp_4)
tr(\gamma_\sigma \gpp_6 \gamma_{\sigma^\prime} \gpp_5)
tr(\gamma_\lambda \gpp_2 \gamma_{\lambda^\prime} \gpp_3),
\end{eqnarray}
where $T_g$ etc. are the color SU(3) group generators, ${\rm g}_{\rm s}$ is
the quark-gluon coupling constant and ${\rm g}_{\rm s}^2 =4\pi \alpha_{\rm s}$,
respectively. The Dirac gamma matrices $\gamma_\lambda$ and $\gamma_\sigma$ in
${\cal M}_{{\rm F}_{*-}}$ are indicated by $\gamma_{\lambda^\prime}$ and 
$\gamma_{\sigma^\prime}$ in ${\cal M}^+_{{\rm F}_{*-}}$.
The calculation of the traces is lengthy and cumbersome, but the 
second Fortran code
is designed to accomplish the task. The Fortran code produces
$$\sum\limits_{\rm {colors}}
tr(T_gT_hT_{h^\prime}T_{g^\prime})
tr(T_hT_{h^\prime})tr(T_gT_{g^\prime})=\frac {4}{3},$$ 
which agrees with the
result derived by hand. Including the average over the spin and color states
of the two initial quarks and the initial antiquark,
the spin- and color-summed squared amplitude is
\begin{eqnarray}
\frac {1}{8} \frac {1}{27} \sum\limits_{\rm {spins,colors}}
\mid {\cal M}_{{\rm F}_{*-}} \mid^2
& = &
\frac {16 {\rm g}_{\rm s}^8}{81}
(  -u_{16}u_{26}u_{34}^2  +u_{16}u_{24}u_{34}u_{35}
      +u_{16}u_{24}u_{34}^2
                   \nonumber   \\
& &
  -u_{16}u_{24}u_{26}u_{34}
  +u_{16}u_{24}^2u_{35}  +u_{16}u_{24}^2u_{34}
                   \nonumber   \\
& &
  -2u_{16}^2u_{24}u_{34} +u_{15}u_{26}u_{34}^2
  -u_{15}u_{24}u_{34}u_{35}
                   \nonumber   \\
& &
  +u_{15}u_{24}u_{34}^2
  +u_{15}u_{24}u_{26}u_{34}  -u_{15}u_{24}^2u_{35}
                   \nonumber   \\
& &
  +u_{15}u_{24}^2u_{34}      -2u_{15}^2u_{24}u_{34}
  -s_{31}u_{15}u_{24}u_{34}
                   \nonumber   \\
& &
  -s_{31}u_{15}u_{24}^2
  -s_{23}u_{16}u_{26}u_{34}  +s_{23}u_{16}u_{24}u_{35}
                   \nonumber   \\
& &
  +2s_{23}u_{16}u_{24}u_{34} +s_{23}u_{16}u_{24}^2
  -s_{23}u_{16}^2u_{34}
                   \nonumber   \\
& &
  -s_{23}u_{16}^2u_{24}
  +s_{23}u_{15}u_{34}^2      +s_{23}u_{15}u_{26}u_{34}
                   \nonumber   \\
& &
  -s_{23}u_{15}u_{24}u_{35}  +2s_{23}u_{15}u_{24}u_{34}
  -s_{23}u_{15}^2u_{34}
                   \nonumber   \\
& &
  -s_{23}u_{15}^2u_{24}
  -s_{23}s_{31}u_{15}u_{24}  +s_{23}^2u_{16}u_{24}
                   \nonumber   \\
& &
  +s_{23}^2u_{15}u_{34}      -s_{12}u_{16}u_{34}^2
  -s_{12}u_{16}u_{24}u_{34}-s_{12}s_{23}u_{16}u_{34} )
                   \nonumber   \\
& &
/[s_{23}(s_{23}+u_{21}+u_{31})(s_{23}-u_{12}-u_{13}+u_{21}+u_{31})]^2,
\end{eqnarray}
which displays the shortest expression among all the individually squared
amplitudes of the diagrams in Figs. 3 and 4,
and the nine independent Lorentz-invariant momentum variables 
$s_{12}=(p_1+p_2)^2$, $s_{23}=(p_2+p_3)^2$, $s_{31}=(p_3+p_1)^2$,
$u_{15}=(p_1-p_5)^2$, $u_{16}=(p_1-p_6)^2$,
$u_{24}=(p_2-p_4)^2$, $u_{26}=(p_2-p_6)^2$,
$u_{34}=(p_3-p_4)^2$ and $u_{35}=(p_3-p_5)^2$.
Interference of amplitudes of different diagrams totals 1004 nonzero terms.
These interference terms are derived with the Fortran codes.
The spin- and color-averaged squared amplitude of every diagram and the
interference terms of different diagrams are used in the transport equation
to study the effects of quark-quark-antiquark elastic scatterings.

\vspace{0.5cm}
\leftline{\bf 4. Transport equation}
\vspace{0.5cm}
Quark matter is assumed to have equal up-quark and down-quark distributions.
The same assumption applies to antiquark matter. The mutual dependence of the
evolution of quark matter and antiquark matter derives from the two-body and
three-body scatterings of quarks and antiquarks. With the assumption that the
quark distribution in quark matter is the same as the antiquark distribution
in antiquark matter, the variation of the up-quark distribution $f_1$ is
described by the transport equation
\begin{eqnarray}
& &
\frac {\partial f_1}{\partial t}
+ \vec {\rm v}_1 \cdot \vec {\nabla}_{\vec {r}} f_1
         \nonumber    \\
& &
= -\frac {{\rm g}_{\rm Q}}{2E_1} \int \frac {d^3p_2}{(2\pi)^32E_2}
\frac {d^3p_3}{(2\pi)^32E_3} \frac {d^3p_4}{(2\pi)^32E_4}
(2\pi)^4 \delta^4(p_1+p_2-p_3-p_4)
         \nonumber    \\
& &
~~~ \times (\frac {1}{2} \mid {\cal M}_{{\rm uu} \to {\rm uu}} \mid^2
+ \mid {\cal M}_{{\rm ud} \to {\rm ud}} \mid^2
+ \mid {\cal M}_{{\rm u\bar u} \to {\rm u\bar u}} \mid^2
+ \mid {\cal M}_{{\rm u\bar d} \to {\rm u\bar d}} \mid^2 )
         \nonumber    \\
& &
~~~ \times [f_1f_2(1-f_3)(1-f_4)-f_3f_4(1-f_1)(1-f_2)]
         \nonumber    \\
& &
~~~ -\frac {{\rm g}_{\rm Q}^2}{2E_1}
\int \frac {d^3p_2}{(2\pi)^32E_2}
\frac {d^3p_3}{(2\pi)^32E_3} \frac {d^3p_4}{(2\pi)^32E_4}
\frac {d^3p_5}{(2\pi)^32E_5} \frac {d^3p_6}{(2\pi)^32E_6}
         \nonumber    \\
& &
~~~ \times (2\pi)^4 \delta^4(p_1+p_2+p_3-p_4-p_5-p_6)
         \nonumber    \\
& &
~~~ \times [\frac {1}{12} \mid {\cal M}_{{\rm uuu} \to {\rm uuu}} \mid^2
+\frac {1}{4} ( \mid {\cal M}_{{\rm uud} \to {\rm uud}} \mid^2
              + \mid {\cal M}_{{\rm udu} \to {\rm udu}} \mid^2 )
+\frac {1}{4} \mid {\cal M}_{{\rm udd} \to {\rm udd}} \mid^2
         \nonumber    \\
& &
~~~ +\frac {1}{2} \mid {\cal M}_{{\rm uu\bar u} \to {\rm uu\bar u}} \mid^2
    +\frac {1}{2} \mid {\cal M}_{{\rm uu\bar d} \to {\rm uu\bar d}} \mid^2
            + \mid {\cal M}_{{\rm ud\bar u} \to {\rm ud\bar u}} \mid^2
            + \mid {\cal M}_{{\rm ud\bar d} \to {\rm ud\bar d}} \mid^2
         \nonumber         \\
& &
~~~ +\frac {1}{4} \mid {\cal M}_{{\rm u\bar {u}\bar u}
                             \to {\rm u\bar {u}\bar u}} \mid^2
    +\frac {1}{2} (
                    \mid {\cal M}_{{\rm u\bar {u}\bar d}
                               \to {\rm u\bar {u}\bar d}} \mid^2
+ \mid {\cal M}_{{\rm u\bar {d}\bar u} \to {\rm u\bar {d}\bar u}} \mid^2 )
+ \frac {1}{4}
  \mid {\cal M}_{{\rm u\bar {d}\bar d} \to {\rm u\bar {d}\bar d}} \mid^2 ]
         \nonumber    \\
& &
~~~ \times [f_1f_2f_3(1-f_4)(1-f_5)(1-f_6)-f_4f_5f_6(1-f_1)(1-f_2)(1-f_3)],
         \nonumber    \\
\end{eqnarray}
where the degeneracy factor ${\rm g}_{\rm Q}=6$
and the velocity of a massless up-quark $v_1=1$.
The distribution function $f_{\rm i}$ is a function of
the position $\vec {r}_{\rm i}$, the momentum $\vec {p}_{\rm i}$ and
the time $t$. The 2-to-2 parton scatterings and the 3-to-3 scatterings
are represented by the first and the second terms on the right-hand side
of the above equation, respectively. Equations for down-quark, up-antiquark
and down-antiquark are written in a similar way. The squared amplitudes for
the 2-to-2 scatterings are the spin- and color-averaged squared amplitudes
of order $\alpha_{\rm s}^2$ obtained in Refs. [20,21].

The squared amplitudes $\mid {\cal M}_{{\rm uuu} \to {\rm uuu}} \mid^2$,
$\mid {\cal M}_{{\rm uud} \to {\rm uud}} \mid^2$,
$\mid {\cal M}_{{\rm udu} \to {\rm udu}} \mid^2$ and
$\mid {\cal M}_{{\rm udd} \to {\rm udd}} \mid^2$ were obtained in
the study of triple-quark elastic scatterings [12].
The calculation of $\mid {\cal M}_{{\rm uu\bar u} \to {\rm uu\bar u}} \mid^2$
requires the individually squared amplitude of every diagram in Figs. 3 and 4
and their interference terms, and the other diagrams arising from
the exchange of the two quarks are also included. The squared amplitude
$\mid {\cal M}_{{\rm uu\bar d} \to {\rm uu\bar d}} \mid^2$ is based on diagrams
$F_-$, $F_+$ and $F_*$ and the diagrams generated from the three diagrams
via the exchange(s) of the two initial quarks and/or the two final quarks;
$\mid {\cal M}_{{\rm ud\bar u} \to {\rm ud\bar u}} \mid^2$ is based on $F_-$,
$F_+$ and $F_*$ and the diagrams generated from all the diagrams in Figs. 3
and 4 via the exchange of the left quark line and the right quark line;
$\mid {\cal M}_{{\rm ud\bar d} \to {\rm ud\bar d}} \mid^2$ is based on all the
diagrams in Figs. 3 and 4 and the two diagrams generated from $F_-$ and $F_+$
via the exchange of both the left quark line and the right quark line. Since
the calculation of $\mid {\cal M}_{{\rm u\bar {u}\bar u}\to {\rm u\bar {u}
\bar u}} \mid^2$ is based on the same set of diagrams as
$\mid {\cal M}_{{\rm uu\bar u} \to {\rm uu\bar u}} \mid^2$,
$\mid {\cal M}_{{\rm u\bar {u}\bar u} \to {\rm u\bar {u}\bar u}} \mid^2$
is obtained from the expression of
$\mid {\cal M}_{{\rm uu\bar u} \to {\rm uu\bar u}} \mid^2$ by the replacements 
$s_{12} \to s_{23}$, $s_{23} \to s_{12}$, $s_{31} \to s_{31}$,
$u_{12} \to u_{32}$, $u_{13} \to u_{31}$, $u_{21} \to u_{23}$,
$u_{23} \to u_{21}$, $u_{31} \to u_{13}$ and $u_{32} \to u_{12}$. Similarly,
$\mid {\cal M}_{{\rm u\bar {u}\bar d} \to {\rm u\bar {u}\bar d}} \mid^2$ is
obtained from
$\mid {\cal M}_{{\rm ud\bar d} \to {\rm ud\bar d}} \mid^2$,
$\mid {\cal M}_{{\rm u\bar {d}\bar u} \to {\rm u\bar {d}\bar u}} \mid^2$ from
$\mid {\cal M}_{{\rm ud\bar u} \to {\rm ud\bar u}} \mid^2$ and
$\mid {\cal M}_{{\rm u\bar {d}\bar d} \to {\rm u\bar {d}\bar d}} \mid^2$ from
$\mid {\cal M}_{{\rm uu\bar d} \to {\rm uu\bar d}} \mid^2$.
These 3-to-3 quark-quark-antiquark scatterings bring more complexity to the
evolution of quark matter than the 3-to-3 quark scatterings alone.

\vspace{0.5cm}
\leftline{\bf 5. Numerical results from the particle sphere technique}
\vspace{0.5cm}
The squared amplitudes
$\mid {\cal M}_{{\rm uuu} \to {\rm uuu}} \mid^2$,
$\mid {\cal M}_{{\rm uud} \to {\rm uud}} \mid^2$,
$\mid {\cal M}_{{\rm udu} \to {\rm udu}} \mid^2$,
$\mid {\cal M}_{{\rm udd} \to {\rm udd}} \mid^2$,
$\mid {\cal M}_{{\rm uu\bar u} \to {\rm uu\bar u}} \mid^2$,
$\mid {\cal M}_{{\rm uu\bar d} \to {\rm uu\bar d}} \mid^2$,
$\mid {\cal M}_{{\rm ud\bar u} \to {\rm ud\bar u}} \mid^2$,
$\mid {\cal M}_{{\rm ud\bar d} \to {\rm ud\bar d}} \mid^2$,
$\mid {\cal M}_{{\rm u\bar {u}\bar u} \to {\rm u\bar {u}\bar u}} \mid^2$,
$\mid {\cal M}_{{\rm u\bar {u}\bar d} \to {\rm u\bar {u}\bar d}} \mid^2$,
$\mid {\cal M}_{{\rm u\bar {d}\bar u} \to {\rm u\bar {d}\bar u}} \mid^2$ and
$\mid {\cal M}_{{\rm u\bar {d}\bar d} \to {\rm u\bar {d}\bar d}} \mid^2$
in Eq. (4) are calculated at $\alpha_{\rm s}=0.5$ and the Coulomb exchange
divergence that is encountered is removed by the use of a screening mass
formulated in Refs. [22-24]. Gluon propagators in Feynman gauge are used in the
squared amplitudes. A screening mass in these propagators leads to some gauge
dependence.
The transport equation is solved until momentum isotropy is established
while up and down quarks and
antiquarks each with number 250 are generated from
the anisotropic parton momentum distribution (1) inside the volume of
$-0.3~{\rm fm}<z<0.3~{\rm fm}$ and $r<6.4$ fm.

When the interaction range is treated as being finite, 
a parton interacts only with surrounding partons. We define a sphere
for a particle which is located at the center of the sphere. Every particle
has a sphere and all the spheres have the same radius. The sphere is called
particle sphere. The amount of particles in a sphere varies with the radius.
The amount of particles also changes from one particle sphere to another. We
denote the maximum amount of particles by $m_{\rm ps}$. The particle sphere 
radius is not the interaction range.  When the sphere radius
equals 1.6 fm, $m_{\rm ps}=75$ is found for 1000 partons generated from the
anisotropic parton momentum distribution. We only allow particles inside the
sphere to scatter. But in the volume defined above this restriction still
produces almost the same number of $n$-parton scatterings as that when no
sphere is defined. Therefore, the restriction can produce accurate result in
$n$-parton scatterings. Since the amount of particles inside a sphere is an
order of magnitude lower than the total number of partons inside the volume,
the operation of searching for scatterings is reduced considerably.
For example, the total number of operations for searching 5-parton scatterings
at time $t=0.2$ fm/$c$ is $1000m_{\rm ps}^4 \approx 3.2 \times 10^{10}$
with the use of particle spheres, which is dramatically smaller than the number
$1000^5=10^{15}$ without the use of particle spheres. Particle spheres form
a basic ingredient in our simulation of parton scatterings.

When a parton traverses inhomogeneous or evolved parton matter, the 
radius-fixed sphere of the parton contains different amounts of partons at
different time. After a multi-parton scattering is finished, a particle sphere
is determined for each final parton. Before or after a scattering event
a parton 
possesses a different particle sphere. However, the particle sphere radius is
always fixed at 1.6 fm independent of time. While parton matter expands, the
amount of partons inside a particle sphere is reduced.

We define a scattering of two partons when the two partons have the closest
distance less than the interaction range of $\sqrt {\sigma_{2 \to 2}/\pi}$.
The 2-to-2 scattering cross section $\sigma_{2 \to 2}$ is
calculated with the spin- and color-averaged squared
amplitudes in Refs. [20,21] which are regulated by the screening mass $\mu_D$,
\begin{equation}
\sigma_{{\rm qq} \to {\rm qq}}=\sigma_{{\rm \bar {q}\bar {q}} \to
{\rm \bar {q}\bar {q}}}
=\frac {{\rm g}_{\rm s}^4}{16\pi s^2} \frac {8}{9} \left[ s
+2 \frac {s(s+2\mu_D^2)^2}{\mu_D^2(s+\mu_D^2)}
+\frac {8}{3}(s+2\mu_D^2)\ln \frac {\mu_D^2}{s+\mu_D^2} \right]
\end{equation}
for the elastic scatterings of two quarks or antiquarks with the same flavor,
\begin{equation}
\sigma_{\rm {qq^\prime} \to \rm {qq^\prime}}
=\sigma_{\rm {\bar {q}\bar {q}^\prime} \to \rm {\bar {q}\bar {q}^\prime}}
=\sigma_{\rm {q\bar {q}^\prime} \to \rm {q\bar {q}^\prime}}
=\frac {{\rm g}_{\rm s}^4}{16\pi s^2} \frac {4}{9} \left[ s
+\frac {2s(s+2\mu_D^2)^2}{\mu_D^2(s+\mu_D^2)}
+2(s+2\mu_D^2)\ln \frac {\mu_D^2}{s+\mu_D^2} \right]
\end{equation}
for the elastic scatterings of quarks and/or antiquarks with different
flavors and
\begin{equation}
\sigma_{\rm {q\bar q} \to \rm {q\bar q}}
=\frac {{\rm g}_{\rm s}^4}{16\pi s^2} \frac {8}{9} \left[
\frac {s(s^2+3\mu_D^2s+3\mu_D^4)}{3(s+2\mu_D^2)^2}
+\frac {s(s+2\mu_D^2)^2}{\mu_D^2(s+\mu_D^2)}
+\frac {2}{3}(s+2\mu_D^2)\ln \frac {\mu_D^2}{s+\mu_D^2} \right]
\end{equation}
for the elastic scatterings of one quark and one antiquark with the same
flavor.

Defining a 3-to-3 scattering event depends on the parton positions and flavors.
If a three-parton scattering occurs, the three partons must be in a sphere
of which the center is at the center-of-mass of the three partons and of
which the radius $r_{\rm hs}$ is [16]
\begin{eqnarray}
\pi r_{\rm hs}^2
& = & \frac {1}{m} \int \frac {d^3p_4}{(2\pi)^32E_4}
\frac {d^3p_5}{(2\pi)^32E_5} \frac {d^3p_6}{(2\pi)^32E_6}
         \nonumber    \\
& &
\times (2\pi)^4 \delta^4(p_1+p_2+p_3-p_4-p_5-p_6)
\mid {\cal M}_{3 \to 3} \mid^2,
\end{eqnarray}
where $m=1$ if $\mid {\cal M}_{3 \to 3} \mid^2 =
\mid {\cal M}_{\rm {ud\bar {u}} \to \rm {ud\bar {u}}} \mid^2$,
$\mid {\cal M}_{\rm {ud\bar {d}} \to {\rm ud\bar {d}}} \mid^2$ or
$\mid {\cal M}_{\rm {u\bar {u}\bar {d}} \to {\rm u\bar {u}\bar {d}}} \mid^2$,
$m=2$ if $\mid {\cal M}_{3 \to 3} \mid^2 =
\mid {\cal M}_{{\rm uu\bar {u}} \to {\rm uu\bar {u}}} \mid^2$ or
$\mid {\cal M}_{{\rm uu\bar {d}} \to {\rm uu\bar {d}}} \mid^2$,
$m=4$ if $\mid {\cal M}_{3 \to 3} \mid^2 =
\mid {\cal M}_{{\rm u\bar {u}\bar {u}} \to {\rm u\bar {u}\bar {u}}} \mid^2$ or
$\mid {\cal M}_{{\rm u\bar {d}\bar {d}} \to {\rm u\bar {d}\bar {d}}} \mid^2$.
Since $\mid {\cal M}_{3 \to 3} \mid^2$ depends on the nine Lorentz-invariant
momentum variables, we replace the phase-space integration by
$du_{14}du_{24}du_{34}du_{15}du_{25}du_{35}$. However, this change is
postponed to the next section.

The use of a finite interaction range in determining 2-to-2 and 3-to-3
scatterings breaks the locality of the transport equation and thus Lorentz
covariance. But the particle subdivision technique [25,26] based on the
transformation $f \to f'=\ell f$ can restore the Lorentz covariance.
At $\ell =20$, a solution of Eq. (4) is taken as the
average of 20 runs of Fortran code starting from
different sets of partons generated
from the distribution (1) at $t=0.2$ fm/$c$.
The solution of the transport equation at the time of the order of 1.75 fm/$c$
is shown in Fig. 5 for various angles by the dotted, dashed and dot-dashed 
curves. The curves overlap and can thus be fitted to the J$\rm \ddot u$ttner
distribution,
\begin{equation}
f(\vec {p})=\frac {\lambda}{{\rm e}^{\mid \vec {p}\mid/T}-\lambda},
\end{equation}
where the temperature of quark matter $T=0.27$ GeV and fugacity
$\lambda=0.31$. We get a thermalization time of 1.55 fm/$c$.

\vspace{0.5cm}
\leftline{\bf 6. Interaction range of three-parton scatterings}
\vspace{0.5cm}
The radius of a sphere defined in Eq. (8)
is Lorentz invariant and depends on $s_{12}$, $s_{23}$ and
$s_{31}$. The integration over $\vec {p}_6$ gives
\begin{equation}
\pi r_{\rm hs}^2
= \frac {1}{m} \int \frac {d^3p_4}{(2\pi)^32E_4}
\frac {d^3p_5}{(2\pi)^32E_5} \frac {2\pi}{2E_6}
\delta (E_1+E_2+E_3-E_4-E_5-E_6)
\mid {\cal M}_{3 \to 3} \mid^2 .
\end{equation}
Since $\mid {\cal M}_{3 \to 3} \mid^2$ is a function of the nine
Lorentz-invariant momentum variables, we replace $d^3p_4$ and $d^3p_5$ with
$du_{14}du_{24}du_{34}du_{15}du_{25}du_{35}$,
\begin{equation}
du_{14}du_{24}du_{34}=J_4dp_{4x}dp_{4y}dp_{4z},
\end{equation}
\begin{equation}
du_{15}du_{25}du_{35}=J_5dp_{5x}dp_{5y}dp_{5z},
\end{equation}
where $u_{14}=(p_1-p_4)^2$, $u_{25}=(p_2-p_5)^2$ and
\begin{equation}
J_4=8E_1E_2E_3\mid (\vec {\rm v}_4 - \vec {\rm v}_1) \cdot
[ (\vec {\rm v}_4 - \vec {\rm v}_2)
\times (\vec {\rm v}_4 - \vec {\rm v}_3) ] \mid ,
\end{equation}
\begin{equation}
J_5=8E_1E_2E_3\mid (\vec {\rm v}_5 - \vec {\rm v}_1) \cdot
[ (\vec {\rm v}_5 - \vec {\rm v}_2)
\times (\vec {\rm v}_5 - \vec {\rm v}_3) ] \mid ,
\end{equation}
where $\vec {\rm v}_{\rm i}$ is the velocity of the ith parton, and
in the center-of-momentum frame the energies of three initial partons in the
3-to-3 scatterings are
\begin{equation}
E_1=\frac {1}{2} \sqrt {\frac {(s_{31}^2+s_s)(s_{12}^2+s_s)}
{s(s_{23}^2+s_s)}},
\end{equation}
\begin{equation}
E_2=\frac {1}{2} \sqrt {\frac {(s_{12}^2+s_s)(s_{23}^2+s_s)}
{s(s_{31}^2+s_s)}},
\end{equation}
\begin{equation}
E_3=\frac {1}{2} \sqrt {\frac {(s_{23}^2+s_s)(s_{31}^2+s_s)}
{s(s_{12}^2+s_s)}},
\end{equation}
with $s_s=s_{12}s_{23}+s_{23}s_{31}+s_{31}s_{12}$ and
the total energy of the three partons 
$E_1 +E_2 +E_3 =
\sqrt {s}=\sqrt {s_{12}+s_{23}+s_{31}}$ is obtained from the definition
$s=(p_1+p_2+p_3)^2$.
Integrating over $u_{25}$ to remove $\delta (E_1+E_2+E_3-E_4-E_5-E_6)$, so
we get
\begin{equation}
\pi r_{\rm hs}^2
 = - \frac {1}{m} \frac {1}{8(2\pi)^5}
\int du_{14} du_{24} du_{34} du_{15} du_{35}
\frac {\mid {\cal M}_{3 \to 3} \mid^2} {E_4 E_5 E_6 J_4 J_5 D} ,
\end{equation}
where in the center-of-momentum frame
the energies of the three final partons are
\begin{equation}
E_4=-\frac {u_{14}+u_{24}+u_{34}}{2\sqrt s},
\end{equation}
\begin{equation}
E_5=-\frac {u_{15}+u_{25}+u_{35}}{2\sqrt s},
\end{equation}
\begin{eqnarray}
E_6 & = & \left\{ E_4^2 +E_5^2
-\frac {2}{d^2W_0}  \left[
\left( W_{11}\frac {u_{15}}{2E_1}+W_{12}\frac {u_{25}}{2E_2}
+W_{31}\frac {u_{35}}{2E_3} \right) \frac {u_{14}}{2E_1}    \right. \right.
                \nonumber       \\
& &
+\left( W_{12}\frac {u_{15}}{2E_1}+W_{22}\frac {u_{25}}{2E_2}
+W_{23}\frac {u_{35}}{2E_3} \right) \frac {u_{24}}{2E_2}
                \nonumber     \\
& &
\left.  \left.
+\left( W_{31}\frac {u_{15}}{2E_1}+W_{23}\frac {u_{25}}{2E_2}
+W_{33}\frac {u_{35}}{2E_3} \right) \frac {u_{34}}{2E_3}    \right]
+O_4O_5
\right\}^{\frac {1}{2}},
\end{eqnarray}
and the absolute value of
the derivative of $E_1+E_2+E_3-E_4-E_5-E_6$ with respect to $u_{25}$ is
\begin{eqnarray}
D  & = & \left| \frac {1}{2\sqrt {s}}
- \frac {1}{2E_6}\left[\frac {u_{15}+u_{25}+u_{35}}{2s}
- \frac {1}{E_2d^2W_0} \left( W_{12}\frac {u_{14}}{2E_1}
+ W_{22} \frac {u_{24}}{2E_2} +W_{23} \frac {u_{34}}{2E_3} \right)
          \right.       \right.
                    \nonumber \\
& &
\left.  \left.
+\frac {O_4}{2E_2O_5} \left( W_{220}\frac {u_{25}}{2E_2} 
+  W_{120}\frac {u_{15}}{2E_1}+W_{230}\frac {u_{35}}{2E_3} \right) \right]
\right| ,
\end{eqnarray}
which is evaluated at $u_{25}$ that is determined by the energy conservation
relation
$E_1+E_2+E_3-E_4-E_5-E_6=0$. Inside Eqs. (21) and (22),
\begin{displaymath}
d=(v_{1x}-v_{2x})(v_{1y}-v_{3y})-(v_{1y}-v_{2y})(v_{1x}-v_{3x}),
\end{displaymath}
\begin{displaymath}
W_0=2(1-\vec {\rm v}_1 \cdot \vec {\rm v}_2)
(1-\vec {\rm v}_2 \cdot \vec {\rm v}_3)
(1-\vec {\rm v}_3 \cdot \vec {\rm v}_1),
\end{displaymath}
\begin{displaymath}
W_{\rm i}=v_{{\rm i}z} (\vec {\rm v}_{\rm j} - \vec {\rm v}_{\rm k})^2 
+ v_{{\rm j}z} (\vec {\rm v}_{\rm j} - \vec {\rm v}_{\rm k}) \cdot
  (\vec {\rm v}_{\rm k} - \vec {\rm v}_{\rm i}) 
+ v_{{\rm k}z} (\vec {\rm v}_{\rm k} - \vec {\rm v}_{\rm j}) \cdot
  (\vec {\rm v}_{\rm j} - \vec {\rm v}_{\rm i}),
\end{displaymath}
\begin{displaymath}
W_{{\rm i}{\rm i}}=W_{\rm i}^2-W_0[(v_{{\rm j}x}-v_{{\rm k}x})^2
+(v_{{\rm j}y}-v_{{\rm k}y})^2],
\end{displaymath}
\begin{displaymath}
W_{{\rm i}{\rm j}}=
\frac {1}{2}(W_{{\rm k}{\rm k}}-W_{{\rm i}{\rm i}}-W_{{\rm j}{\rm j}})
\end{displaymath}
\begin{displaymath}
W_{{\rm i}{\rm i}0}=W_{{\rm i}{\rm i}}
+W_0(v_{{\rm j}x}v_{{\rm k}y}-v_{{\rm j}y}v_{{\rm k}x})^2,
\end{displaymath}
\begin{displaymath}
W_{{\rm i}{\rm j}0}=W_{{\rm i}{\rm j}}
+W_0(v_{{\rm j}x}v_{{\rm k}y}-v_{{\rm j}y}v_{{\rm k}x})
(v_{{\rm k}x}v_{{\rm i}y}-v_{{\rm k}y}v_{{\rm i}x}),
\end{displaymath}
where the subscripts ${\rm i},{\rm j}$ and ${\rm k}$ allow the three
cases: ${\rm i}=1$, ${\rm j}=2$, ${\rm k}=3$; 
${\rm i}=2$, ${\rm j}=3$, ${\rm k}=1$;
${\rm i}=3$, ${\rm j}=1$, ${\rm k}=2$; and 
\begin{eqnarray}
O_{\rm n} & = & \left[
W_{110}\left( \frac {u_{1{\rm n}}}{2E_1} \right)^2
+W_{220}\left( \frac {u_{2{\rm n}}}{2E_2} \right)^2
+W_{330}\left( \frac {u_{3{\rm n}}}{2E_3} \right)^2     \right.
                    \nonumber     \\
& &
\left.
+2W_{120} \frac {u_{1{\rm n}}}{2E_1} \frac {u_{2{\rm n}}}{2E_2}
+2W_{230} \frac {u_{2{\rm n}}}{2E_2} \frac {u_{3{\rm n}}}{2E_3}
+2W_{310} \frac {u_{1{\rm n}}}{2E_1} \frac {u_{3{\rm n}}}{2E_3}
\right]^{\frac {1}{2}},                \nonumber
\end{eqnarray}
with ${\rm n}=4,5$. The quantity
$u_{25}$ determined by energy conservation gives
\begin{equation}
u_{26}=-s_{12}-s_{23}-u_{24}-u_{25}
\end{equation}
that appears in $\mid {\cal M}_{3 \to 3} \mid^2$.
Replacing $u_{14}$ with $u_{16}$ in
$u_{14}=-(s_{12}+s_{31}+u_{15}+u_{16})$, we obtain
\begin{equation}
\pi r_{\rm hs}^2
 = - \frac {1}{m} \frac {1}{8(2\pi)^5}
\int du_{15} du_{16} du_{24} du_{34} du_{35}
\frac {\mid {\cal M}_{3 \to 3} \mid^2} {E_4 E_5 E_6 J_4 J_5 D}.
\end{equation}
The variables $u_{15}$, $u_{16}$, $u_{24}$, $u_{34}$ and $u_{35}$
have different minima:
\begin{equation}
u_{1 \rm min}= 
-\sqrt {(s_{31}^2+s_s)(s_{12}^2+s_s)/(s_{23}^2+s_s)}=-(s_{31}+s_{12}),
\end{equation}
for $u_{15}$ and $u_{16}$,
\begin{equation}
u_{2 \rm min}=
-\sqrt {(s_{12}^2+s_s)(s_{23}^2+s_s)/(s_{31}^2+s_s)}=-(s_{12}+s_{23}),
\end{equation}
for $u_{24}$ and
\begin{equation}
u_{3 \rm min}=
-\sqrt {(s_{23}^2+s_s)(s_{31}^2+s_s)/(s_{12}^2+s_s)}=-(s_{23}+s_{31}),
\end{equation}
for $u_{34}$ and $u_{35}$. In the center-of-momentum frame $E_4 +E_5 +E_6 =
\sqrt {s}$ and $\vec {p}_4+\vec {p}_5+\vec {p}_6=0$. While the three vectors
$\vec {p}_4$, $\vec {p}_5$ and $\vec {p}_6$ are parallel, 
$\mid \vec {p}_4 \mid$, $\mid \vec {p}_5 \mid$ or $\mid \vec {p}_6 \mid$ takes 
its maximum $\sqrt {s}/2$. If, for example, $\vec {p}_1$ and $\vec {p}_5$
point to opposite directions and $\vec {p}_1$, $\vec {p}_4$ and $\vec {p}_6$
point to the same direction, $u_{15}$ by its definition reaches 
$-2E_1\sqrt {s}$, i.e. its minimum. Hence, when one of
variables $u_{15}$, $u_{16}$, $u_{24}$, $u_{34}$ and $u_{35}$ takes its 
minimum, the three vectors $\vec {p}_4$, $\vec {p}_5$ 
and $\vec {p}_6$ and one of
$\vec {p}_1$, $\vec {p}_2$ and $\vec {p}_3$  are parallel. The minima set
useful bounds to the Lorentz-invariant momentum variables by
\begin{equation}
u_{14}+u_{15}+u_{16}=u_{1 \rm min}
\end{equation}
\begin{equation}
u_{24}+u_{25}+u_{26}=u_{2 \rm min}
\end{equation}
\begin{equation}
u_{34}+u_{35}+u_{36}=u_{3 \rm min}
\end{equation}
with $u_{36}=(p_3-p_6)^2$. Eq. (28) means that $u_{15}$ and $u_{16}$ cannot 
take the minimum $u_{1 \rm min}$ at the same time. If $u_{15}=u_{1 \rm min}$,
$u_{14}=u_{16}=0$. Eq. (28) is a plane equation in three-dimensional space.
The distances of the $u_{14}$-intercept, $u_{15}$-intercept and 
$u_{16}$-intercept of the 
plane to the origin are the same, $-u_{1 \rm min}$. Since $u_{14}<0$, 
$u_{15}<0$ and $u_{16}<0$, the regions of the three variables are  
the interior of the triangle in the plane of which the three sides connect
the $u_{14}$-intercept, $u_{15}$-intercept and $u_{16}$-intercept. 
The triangle is an equilateral triangle that has a side length of 
$-\sqrt {2}u_{1 \rm min}$. Similar discussion applies
to Eqs. (29) and (30). The three planes indicated by Eqs. (28)-(30) are 
parallel.

\vspace{0.5cm}
\leftline{\bf 7. Summary}
\vspace{0.5cm}
We have derived squared amplitudes for $qq\bar {q} \to qq\bar q$ and
$q\bar {q}\bar {q} \to q\bar {q}\bar q$ at the tree level. Feynman diagrams
depicted in Figs. 3 and 4 show the 3-to-3 scatterings which lead to the
interplay of quark matter and antiquark matter. These squared amplitudes form
new contributions in transport equations for quarks and antiquarks. We have
presented the particle sphere technique to solve the transport equations
while the formula for the interaction range of three partons is obtained by
an integration over the Lorentz-invariant momentum variables in the equilateral
triangle regions. It is shown by 
the solutions of the transport equations that the thermalization time of quark
matter and antiquark matter is of the order of about 1.55 fm/$c$. This
reduced thermalization time is an effect of quark-quark-antiquark and
quark-antiquark-antiquark elastic scatterings that suggests the importance
of multi-parton scatterings at high parton number density. Another interesting 
effect of three-body elastic scattering is shown on heavy quark momentum 
degradation in quark-gluon plasma [27].

\vspace{0.5cm}
\leftline{\bf Acknowledgements}
\vspace{0.5cm}
This work was supported in part by the National Natural Science Foundation of
China under Grant No. 10675079. X.-M. thanks C.M. Ko for bringing his 
interesting work to the author's attention during quark matter 2006 conference 
at Shanghai.

\newpage
\leftline{\bf References}
\vskip 14pt
\leftline{[1]STAR Collaboration, J. Adams, Nucl. Phys. {\bf A757} (2005) 102.}
\leftline{[2]PHENIX Collaboration, K. Adcox, Nucl. Phys. {\bf A757} (2005)
184.}
\leftline{[3]E. Shuryak, Phys. Rev. Lett. {\bf 68} (1992) 3270.}
\leftline{[4]K. Geiger, Phys. Rev. {\bf D46} (1992) 4965;}
\leftline{~~~K. Geiger, Phys. Rev. {\bf D46} (1992) 4986.}
\leftline{[5]G.R. Shin, B. M$\rm \ddot u$ller, J. Phys. {\bf G29} (2003)
2485.}
\leftline{[6]Z. Xu, C. Greiner, Phys. Rev. {\bf C71} (2005) 064901.}
\leftline{[7]S.M.H. Wong, Phys. Rev. {\bf C54} (1996) 2588.}
\leftline{~~~G.C. Nayak, A. Dumitru, L. McLerran, W. Greiner, Nucl. Phys.
{\bf A687} (2001) 457.}
\leftline{[8]P.F. Kolb, P. Huovinen, U. Heinz, H. Heiselberg, Phys. Lett.
{\bf B500} (2001) 232.}
\leftline{~~~P. Huovinen, Nucl. Phys. {\bf A715} (2003) 299c.}
\leftline{[9]D. Teaney, J. Lauret, E.V. Shuryak, nucl-th/0110037.}
\leftline{~~~E.V. Shuryak, Nucl. Phys. {\bf A715} (2003) 289c.}
\leftline{[10]T. Hirano, Phys. Rev. {\bf C65} (2001) 011901.}
\leftline{~~~~K. Morita, S. Muroya, C. Nonaka, T. Hirano, Phys. Rev.
{\bf C66} (2002) 054904.}
\leftline{[11]K.J. Eskola, H. Niemi, P.V. Ruuskanen, S.S.
R$\rm \ddot a$s$\rm \ddot a$nen, Phys. Lett. {\bf B566} (2003) 187;}
\leftline{~~~~K.J. Eskola, H. Niemi, P.V. Ruuskanen, S.S.
R$\rm \ddot a$s$\rm \ddot a$nen, Nucl. Phys. {\bf A715} (2003) 561c.}
\leftline{[12]X.-M. Xu, R. Peng, H.J. Weber, Phys. Lett. {\bf B629} (2005) 68.}
\leftline{[13]K.J. Eskola, K. Kajantie, K. Tuominen, Phys. Lett. {\bf B497}
(2001) 39.}
\leftline{[14]M. Gyulassy, P. L$\rm \acute e$vai, I. Vitev, Nucl. Phys.
{\bf B594} (2001) 371;}
\leftline{~~~~M. Gyulassy, I. Vitev, X.-N. Wang, P. Huovinen,
Phys. Lett. {\bf B526} (2002) 301.}
\leftline{[15]F. Cooper, E. Mottola, G.C. Nayak, Phys. Lett. {\bf B555}
(2003) 181.}
\leftline{[16]X.-M. Xu, Y. Sun, A.-Q. Chen, L. Zheng, Nucl. Phys.
{\bf A744} (2004) 347.}
\leftline{[17]X.-N. Wang, M. Gyulassy, Phys. Rev. {\bf D44} (1991) 3501;}
\leftline{~~~~M. Gyulassy, X.-N. Wang, Comput. Phys. Commun. {\bf 83}
(1994) 307;}
\leftline{~~~~X.-N. Wang, Phys. Rep. {\bf 280} (1997) 287.}
\leftline{[18]P. L$\rm \acute e$vai, B. M$\rm \ddot u$ller, X.-N. Wang, Phys.
Rev. {\bf C51} (1995) 3326.}
\leftline{[19]R.D. Field, {\it Applications of Perturbative QCD},
Addison-Wesley, Redwood City, 1989.}
\leftline{[20]R. Cutler, D. Sivers, Phys. Rev. {\bf D17} (1978) 196.}
\leftline{[21]B.L. Combridge, J. Kripfganz, J. Ranft, Phys. Lett.
{\bf B70} (1977) 234.}
\leftline{[22]T.S. Bir$\rm \acute o$, B. M$\rm \ddot u$ller,
X.-N. Wang, Phys. Lett. {\bf B283} (1992) 171.}
\leftline{[23]K.J. Eskola, B. M$\rm \ddot u$ller,
X.-N. Wang, Phys. Lett. {\bf B374} (1996) 20.}
\leftline{[24]S.A. Bass, B. M$\rm \ddot u$ller,
D.K. Srivastava, Phys. Lett. {\bf B551} (2003) 277.}
\leftline{[25]B. Zhang, M. Gyulassy, Y. Pang, Phys. Rev. {\bf C58} (1998)
1175.}
\leftline{[26]D. Moln$\rm \acute a$r, M. Gyulassy, Nucl. Phys.
{\bf A697} (2002) 495.}
\leftline{[27]W. Liu, C.M. Ko, nucl-th/0603004.}

\newpage
\begin{figure}[t]
  \begin{center}
    \includegraphics[width=0.7\textwidth,angle=0]{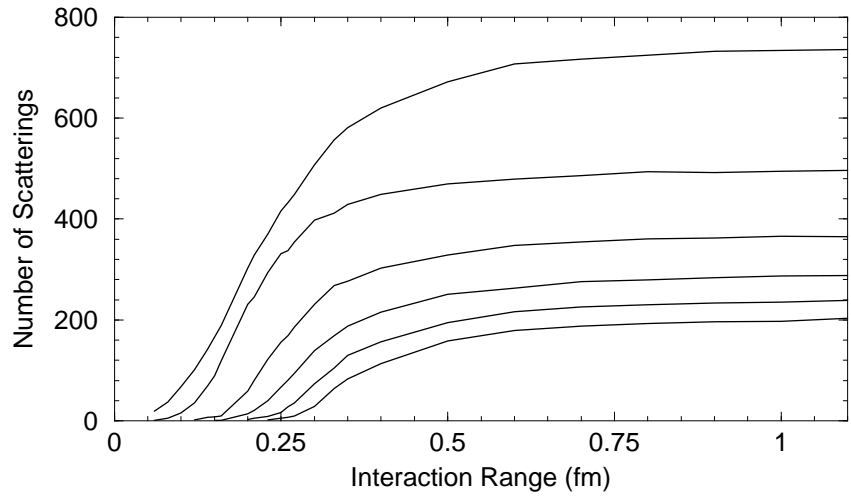}
  \end{center}
\caption{Curves from top to bottom show the numbers of $n$-parton scatterings
with $n=2, \cdot \cdot \cdot , 7$, respectively.}
\label{fig1}
\end{figure}

\newpage
\begin{figure}[t]
  \begin{center}
    \includegraphics[width=0.7\textwidth,angle=0]{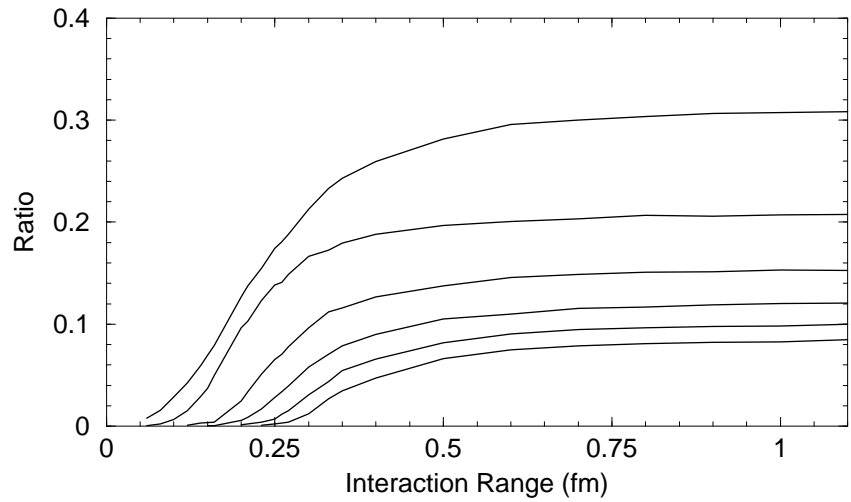}
  \end{center}
\caption{Curves from top to bottom show the ratio $R_n$
with $n=2, \cdot \cdot \cdot , 7$, respectively.}
\label{fig2}
\end{figure}

\newpage
\begin{figure}[t]
  \begin{center}
    \leavevmode
    \includegraphics[width=42mm,height=65mm,angle=0]{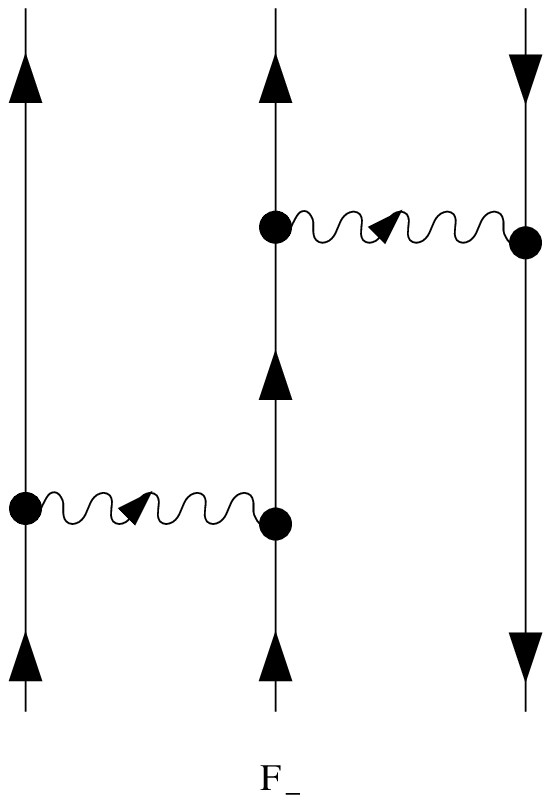}
      \hspace{1.2cm}
    \includegraphics[width=42mm,height=65mm,angle=0]{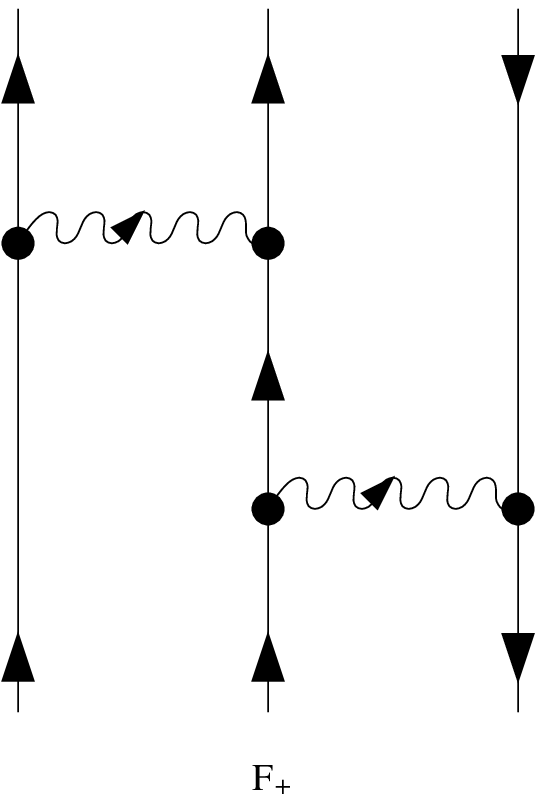}
      \hspace{1.2cm}
    \includegraphics[width=42mm,height=65mm,angle=0]{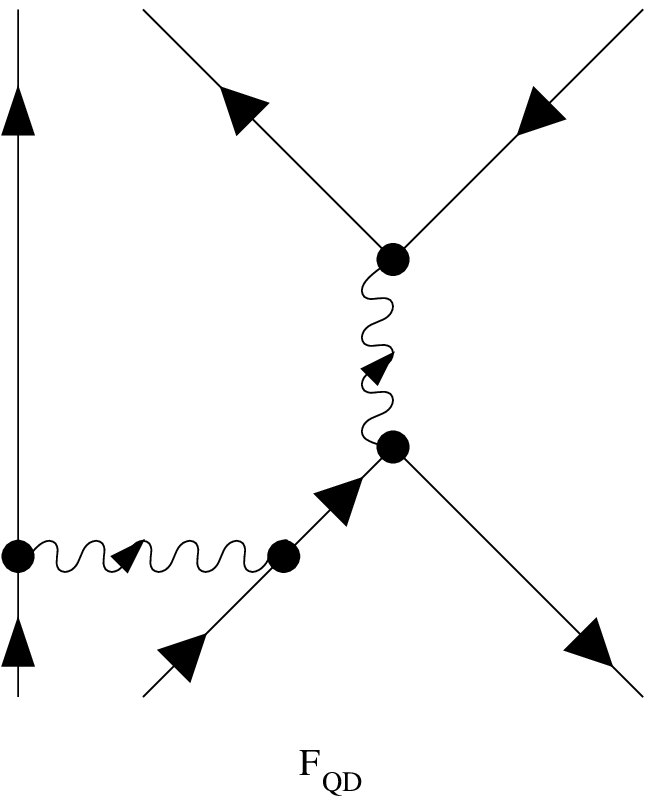}
      \vskip 26pt
    \includegraphics[width=42mm,height=65mm,angle=0]{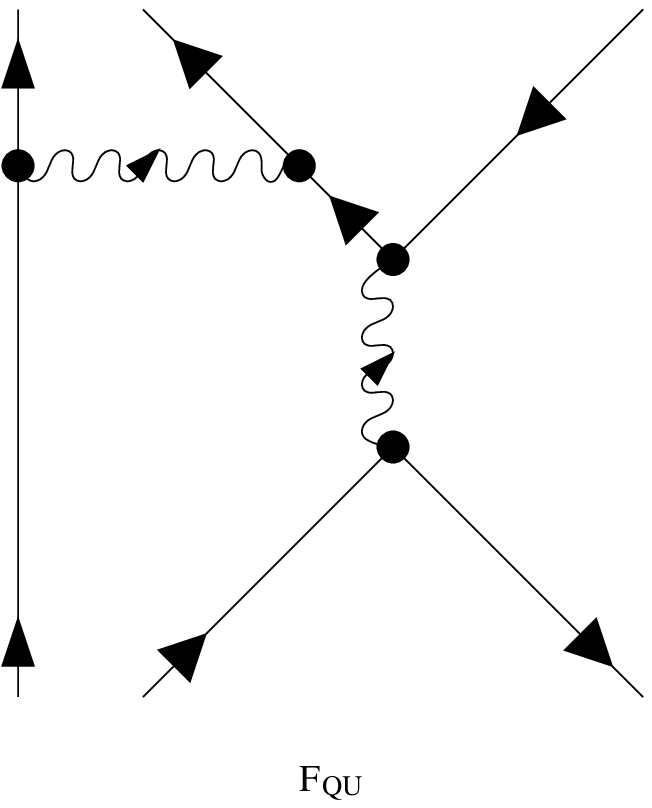}
      \hspace{1.2cm}
    \includegraphics[width=42mm,height=65mm,angle=0]{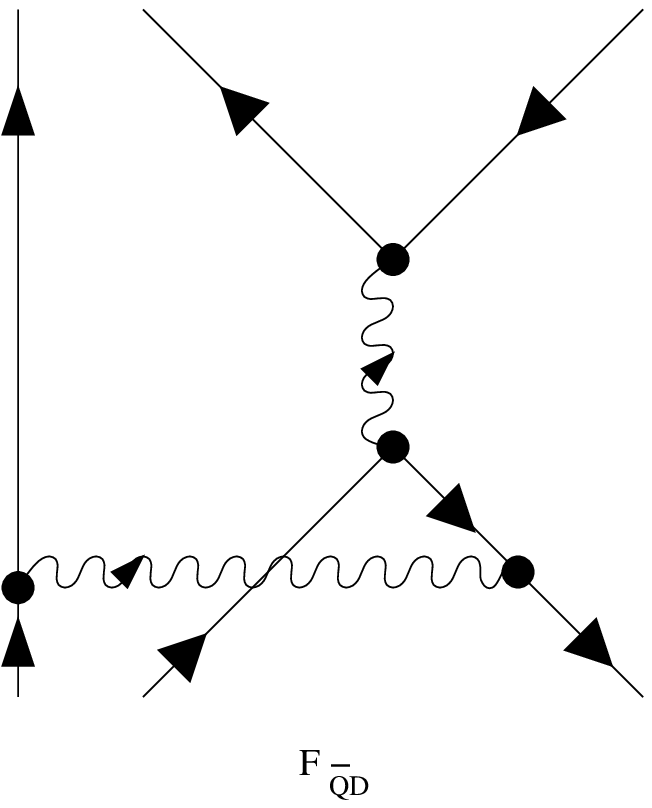}
      \hspace{1.2cm}
    \includegraphics[width=42mm,height=65mm,angle=0]{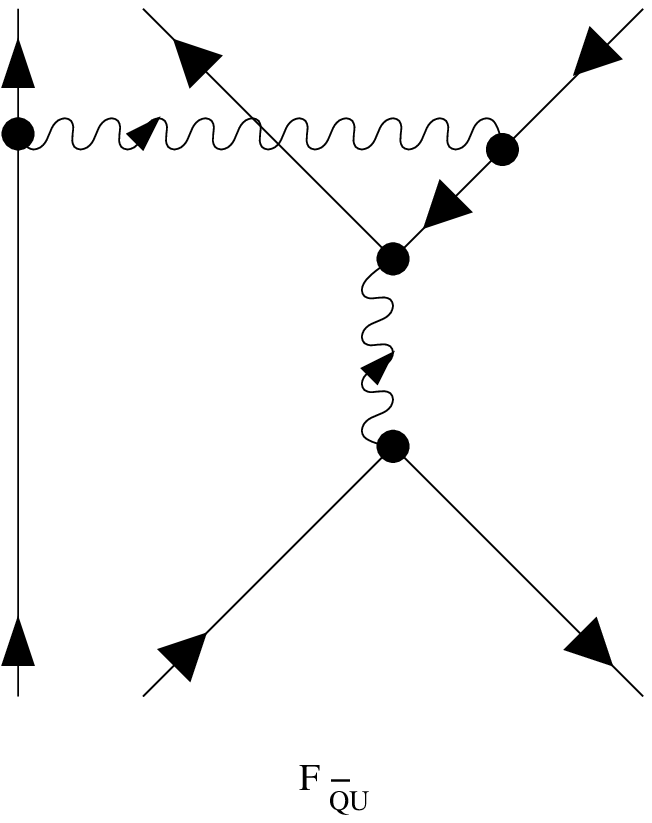}
      \vskip 26pt
    \includegraphics[width=42mm,height=65mm,angle=0]{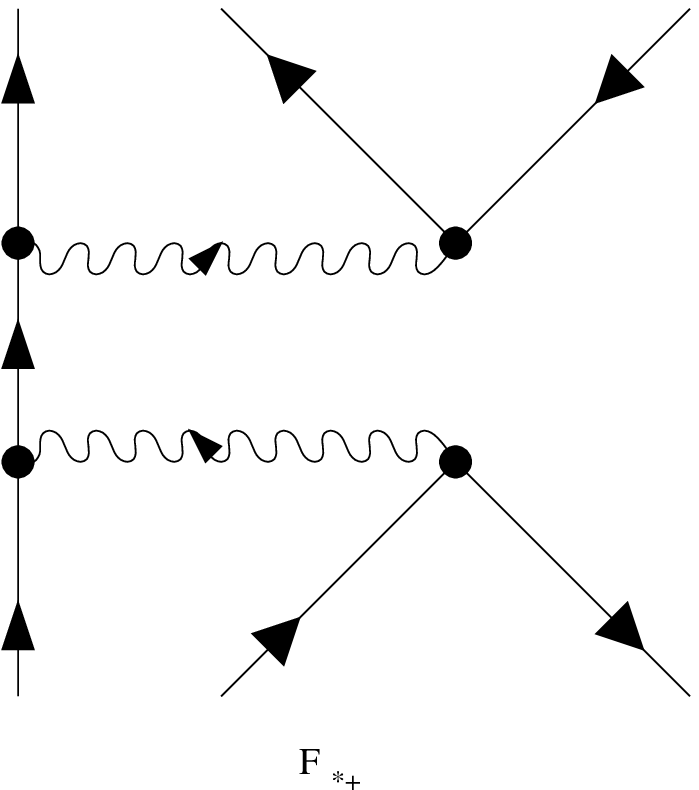}
      \hspace{1.2cm}
    \includegraphics[width=42mm,height=65mm,angle=0]{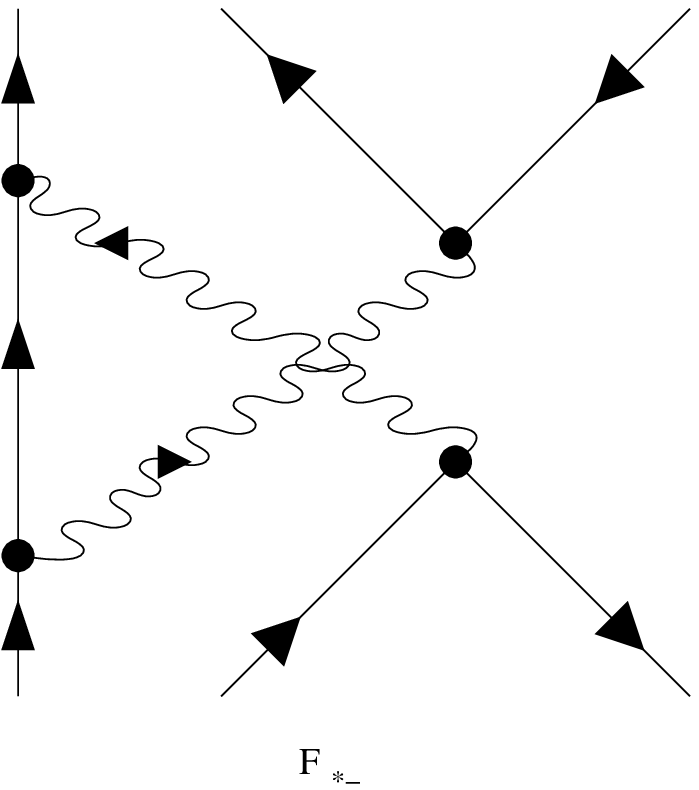}
      \hspace{1.2cm}
  \end{center}
\caption{Two-gluon-exchange induced scatterings of two quarks and one
antiquark.}
\label{fig3}
\end{figure}

\newpage
\begin{figure}[t]
  \begin{center}
    \includegraphics[width=42mm,height=65mm,angle=0]{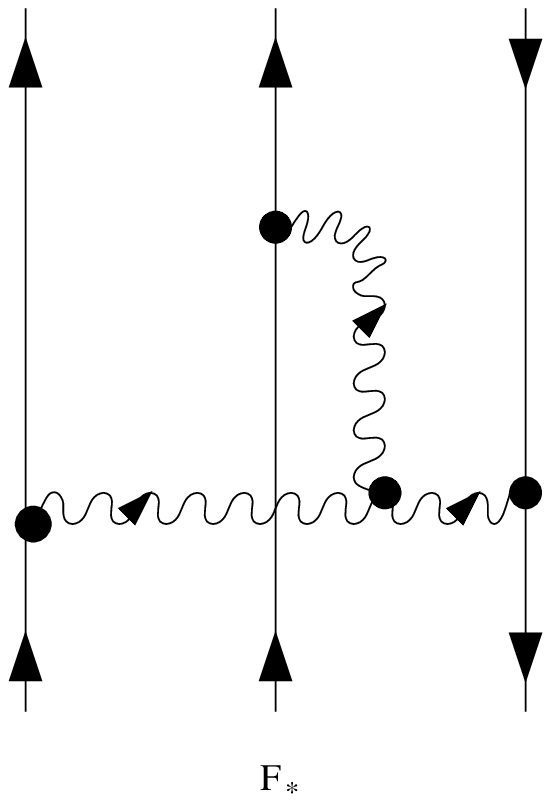}
      \hspace{1.2cm}
    \includegraphics[width=42mm,height=65mm,angle=0]{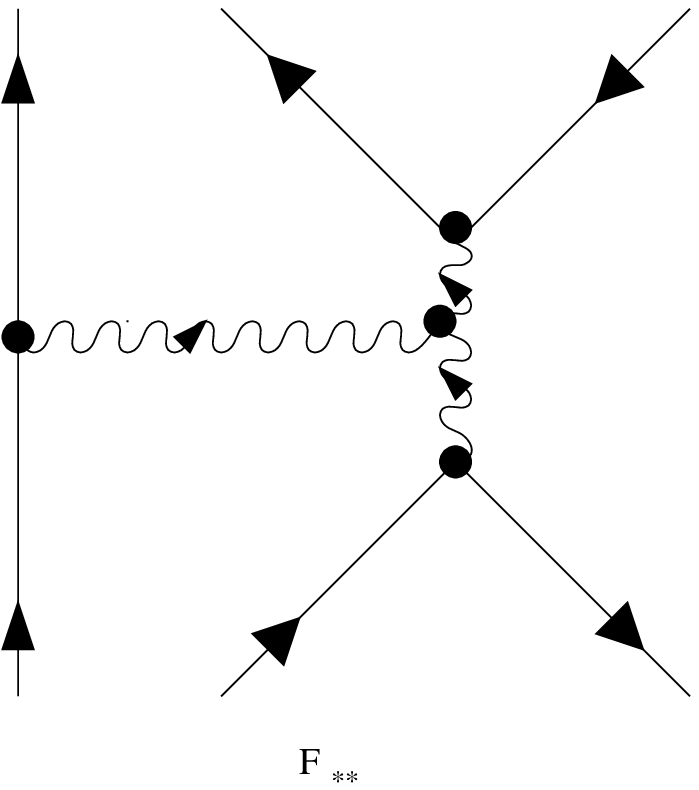}
      \hspace{1.2cm}
  \end{center}
\caption{Triple-gluon coupling in quark-quark-antiquark scatterings.}
\label{fig4}
\end{figure}

\newpage
\begin{figure}[t]
  \begin{center}
    \includegraphics[width=0.7\textwidth,angle=0]{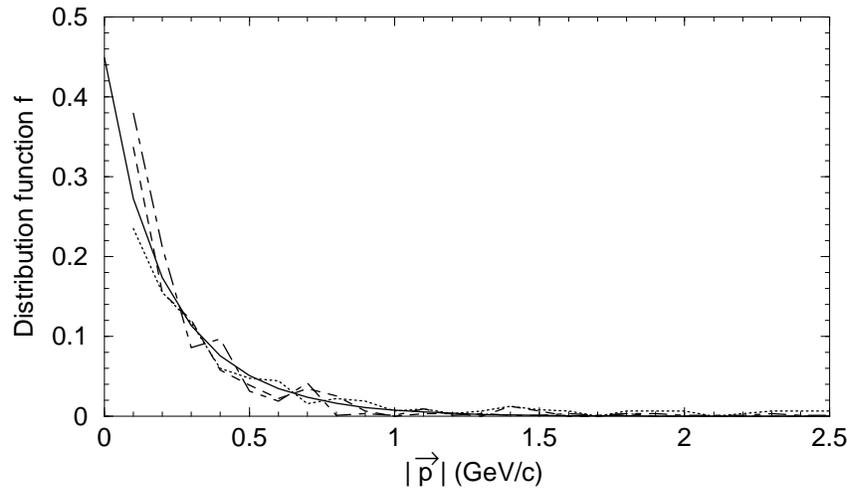}
  \end{center}
\caption{Quark distribution functions versus momentum in different
directions while  quark matter arrives at thermal equilibrium.
The dotted, dashed and dot-dashed curves correspond to the angles
relative to one incoming beam direction
$\theta =0^{\rm o}, 45^{\rm o}, 90^{\rm o}$, respectively.
The solid curve represents the thermal distribution function.}
\label{fig5}
\end{figure}

\end{document}